\documentclass[a4paper]{article}

\usepackage{multicol,multirow,graphicx,graphics,fancyhdr}
\usepackage{amsmath,amsfonts,amssymb,bm,upgreek,mathrsfs,mathcomp,float,subcaption}
\usepackage[pagewise,switch,columnwise]{lineno}
\usepackage[compress,nospace]{cite}
\usepackage[dvipsnames]{xcolor}
\usepackage{CPL-20232}

\begin{document}
\vspace* {-4mm} 
\begin{center}
%-------------------------Title----------------------------

\large\bf{\boldmath{Coupling phase enabled level transitions in pseudo-Hermitian magnon-polariton systems}}
%------------------------Footnote--------------------------
\footnotetext{
\hspace*{-5.4mm}$^{\dagger}$These authors contributed equally\\
\hspace*{-0.5mm}$^{*}$Corresponding authors. Email: shironglin@gbu.edu.cn}
\\[5mm]
%------------------------Authors----------------------------
\normalsize \rm{}Xin Huang$^{1,2\dagger}$, Jingyu Liu$^{1,2\dagger}$, and Shirong Lin$^{1,2*}$
%----------------------COM. or University-------------------
\\[3mm]\small\sl $^{1}$School of Physical Sciences, Great Bay University, Dongguan 523000, China

$^{2}$Great Bay Institute for Advanced Study, Dongguan 523000, China\\[4mm]

%------------------------Received date----------------------

\normalsize\rm{} \hspace{2cm} 
\end{center}

%----------------------Abstract and PACS--------------------
\vskip 1.5mm

\small{\narrower While cavity-magnon hybridization offers intriguing physics, its practical implementation is hindered by intrinsic damping in both cavity and magnon modes, leading to short coherence times and constrained applications. Recently, with the emergence of tunable external gain at the macroscopic scale, the research focus has shifted from purely lossy systems to gain-loss balanced non-Hermitian systems. Here, we propose a pseudo-Hermitian model with two magnon and two cavity modes coupled via phase-dependent interaction. We link the energy spectrum to phase transitions, observing exceptional points when pseudo-Hermitian symmetry breaks.  We also observed level attraction and level repulsion. The former corresponds to four phase transitions and manifests as a double Z-shaped energy spectrum, the latter corresponds to two phase transitions, with the repulsive gap depending on the coupling phase.  In the phase diagram defined by non-Hermiticity and coupling phase, we reveal a distinctive correspondence: pseudo-Hermitian symmetry breaking is intrinsically linked to coupling mode transitions, enabling new strategies for controlling hybrid quantum states in spintronic systems.

\par}\vskip 3mm
\normalsize\noindent{\narrower{\nolinkurl{}{}}

\par}
\vskip 5mm
%-------------------TEXT TEXT TEXT TEXT---------------------
\begin{multicols}{2}
{\textit{Introduction}}\ \ Controlling light-matter interactions is a key research focus.\ucite{1,2,3,4,5,6,7,8,57} Magnons,\ucite{55,56,58,59} collective excitations in ferromagnets like yttrium iron garnet (YIG),\ucite{12} interact strongly with microwave photons, forming cavity magnon polaritons (CMPs).\ucite{13,14,15} These quasiparticles offer long coherence times and high tunability, enabling quantum manipulation.\ucite{16,17,18,19} CMP systems exhibit diverse phenomena, including coherent perfect absorption,\ucite{5} dark states,\ucite{3} multistability,\ucite{6,21} magnon quintuplet states,\ucite{22} single-spin detection,\ucite{23} and spin-current control.\ucite{24} Beyond coherent coupling, dissipative coupling shows unique properties like level attraction and damping repulsion.\ucite{26,27}  It enables nonreciprocal microwave propagation,\ucite{29,30} unidirectional invisibility,\ucite{29} and stable Bell states.\ucite{31}

On the other hand, pseudo-Hermitian systems are those that do not satisfy traditional Hermitian conditions, but still maintain a real eigenvalue spectrum. The variation of system parameters may lead to the breakdown of pseudo-Hermiticity, causing the eigenvalue spectrum to transform from real to complex. It is usually accompanied by the appearance of exceptional points (EPs), which refers to the simultaneous degeneracy of both eigenvalues and eigenstates.\ucite{32,33,34}

In previous studies, researchers have regulated the hybridization of CMPs through various methods, including designing multiple types of resonant cavities,\ucite{35} using different materials,\ucite{37} adjusting the sample positioning within the cavities.\ucite{39} The early CMP systems have short coherence times due to the inherent damping of both cavity modes and magnon modes, limiting the application of CMP systems. However, by continuously pumping to compensate for the decay and constructing a non-Hermitian system,\ucite{28,42,43,44,45,60} not only can the coherence times of CMPs be extended,\ucite{5,17} but the pseudo-Hermitian system can also maintain a real eigenvalue spectrum.

We investigate a pseudo-Hermitian cavity-magnon system coupling two cavity and two magnon modes. We link the energy spectrum
to phase transition, revealing EPs  when pseudo-Hermiticity breaks in pseudo-Hermitian phase diagram. Key parameters are non-Hermiticity and coupling phase. At zero phase, dissipative coupling induces level attraction with a characteristic double Z-shaped spectrum, corresponding to four phase transitions. With nonzero phases, coherent coupling also emerges, producing level repulsion with two transitions and phase-dependent repulsive gaps. In the pseudo-Hermitian phase diagram defined by both non-Hermiticity and the coupling phase, pseudo-Hermiticity breaking is uniquely mapped to transitions between coupling modes, offering novel control strategies for cavity-magnon systems.

{\textit{Model}}\ \ Two YIG spheres, with gain and loss, couple to two microwave cavity modes (Fig. 1(a)). Both have the same magnon frequency, tuned by the magnetic field \( \text{\textbf{B}}_\textbf{0} \). For the magnon modes, we consider small amplitude excitations and employ the Holstein-Primakoff transformation to describe these bosonic excitations.\ucite{46} The total Hamiltonian is given by \( \mathcal{H}_{\mathrm{total}} = \mathcal{H}_{\mathrm{free}} + \mathcal{H}_{\mathrm{I}} \), The free Hamiltonian is described by\begin{equation}
\begin{aligned}
\mathcal{H}_{\mathrm{free}} = & \hbar \omega_{c,1} c_{1}^{\dagger} c_{1} + \hbar \omega_{c,2} c_{2}^{\dagger} c_{2} \\
& + \hbar (\omega_{m} - i \gamma) m_1^{\dagger} m_1 + \hbar (\omega_{m} + i \gamma) m_2^{\dagger} m_2,
\end{aligned}
\end{equation}\begin{figure}[H]
  \centering
  \begin{subfigure}[b]{0.4\textwidth}
    \includegraphics[width=\textwidth, height=0.59\textwidth]{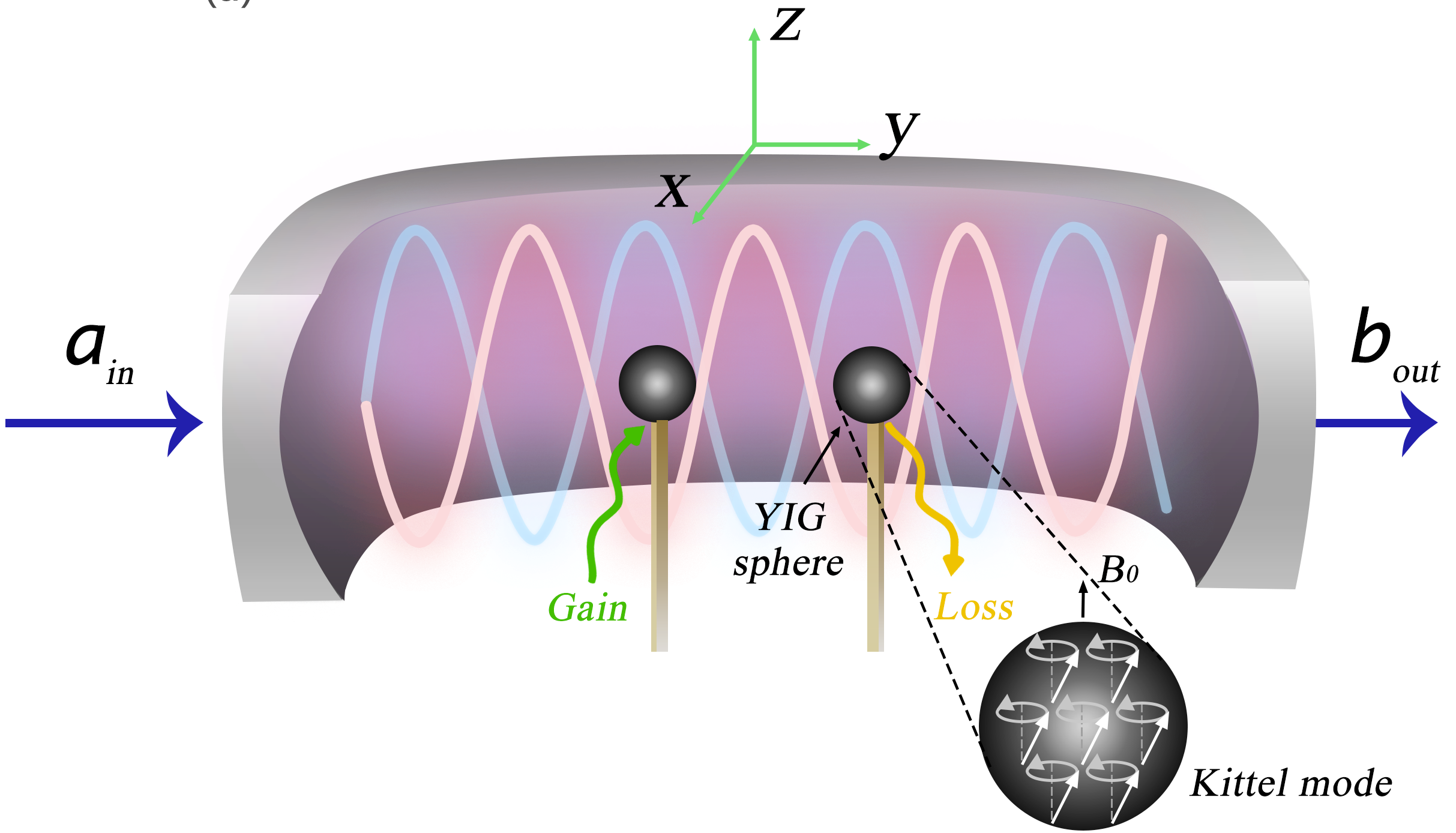}
    \caption{}
    \label{fig:sub1}
  \end{subfigure}
  \begin{subfigure}[b]{0.45\textwidth}
    \includegraphics[width=\textwidth, height=0.5\textwidth]{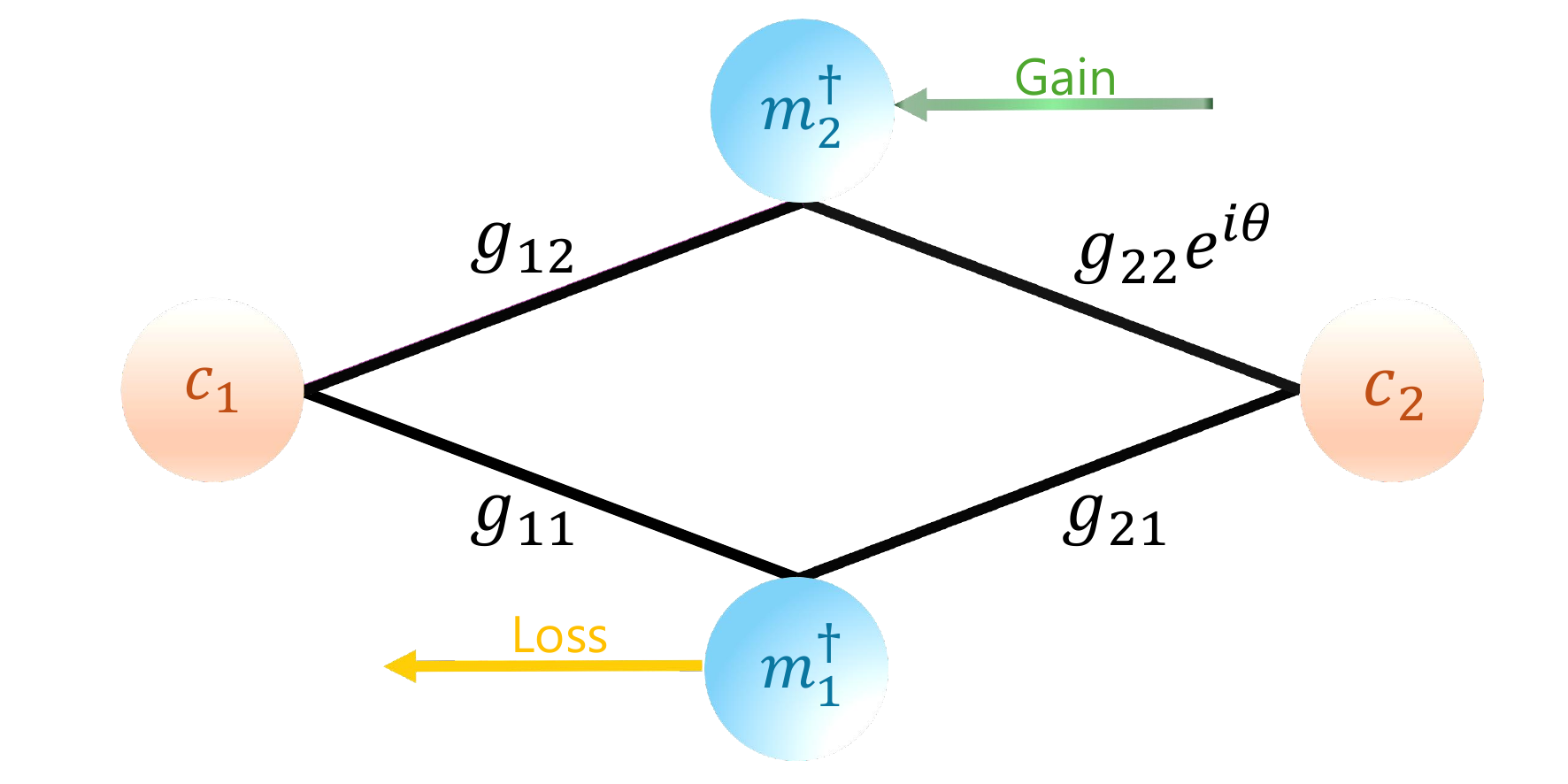}
    \caption{}
    \label{fig:sub2}
  \end{subfigure}  \vspace{-5pt}
  \caption{(a) Coupling model diagram.
(b) Schematic showing interactions and net phase difference from the loop-coupled architecture.}
  \label{fig:sub3}
\end{figure}
\noindent \( c_k \) and \( m_k \) are the boson  annihilation operators for the \( k \)-th cavity and magnon mode \( (k = 1, 2) \). Here,  $-i\gamma$ and $+i\gamma$ 
 represent energy loss and gain for the first and second YIG spheres, respectively. Non-Hermitian systems with external gain and loss have been studied, especially for PT- or anti-PT-symmetric frameworks.\ucite{5,47,49} Within the rotating wave approximation, the interaction Hamiltonian
is \begin{equation}
\begin{aligned}
\mathcal{H}_I &= \hbar g_{11} e^{i\varphi_{11}} c_1 m_1^\dagger + \hbar g_{12} e^{i\varphi_{12}} c_1 m_2^\dagger \\
     &+ \hbar g_{21} e^{i\varphi_{21}} c_2 m_1^\dagger + \hbar g_{22} e^{i\varphi_{22}} c_2 m_2^\dagger + \mathrm{h.c},
\end{aligned}
\end{equation}
where \( g_{k,l} \, (k,l = 1,2) \) denote the coupling strength, and \(\varphi_{k,l} \) represent the coupling phase from the Zeeman interaction between the magnon and cavity mode.\ucite{50,51} These coupling phases, arising from RF magnetic field orientations in cavity modes interacting with magnetic samples, are crucial when two cavity modes couple with two magnon modes, enabling synthetic gauge fields and dark modes.\ucite{52,53} We apply the unitary transformations \( U_1 = e^{i\varphi_{11} c_1^{\dagger} c_1} \), \( U_2 = e^{i\varphi_{21} c_2^{\dagger} c_2} \) and \( U_3 = e^{i (\varphi_{12} - \varphi_{11}) m_2 m_2^{\dagger}} \) to consolidate the four coupling phases in Eq. (3) into a single gauge-invariant phase term. The simplified Hamiltonian is as follows, \begin{equation}
\begin{aligned}
\mathcal{H}_I &= hg_{11} c_1 m_1^\dagger + hg_{12} c_1 m_2^\dagger \\
     &\quad + hg_{21} c_2 m_1^\dagger + hg_{22} e^{i \theta} c_2 m_2^\dagger + \mathrm{h.c.},
\end{aligned}
\end{equation} 
where \( \theta = \varphi_{22} - \varphi_{21} - \varphi_{12} + \varphi_{11} \) encapsulates the net phase
difference arising from the loop-coupled architecture.
We assume all magnon-photon couplings have identical strength \( g \). In Appendix A, we prove that the total Hamiltonian satisfies \( \mathcal{H}^\dagger = \eta \mathcal{H} \eta^{-1} \), where \( \mathbf{\eta} \) is invertible. Thus, the Hamiltonian is pseudo-Hermitian, meaning it can have a real spectrum. We diagonalize the total Hamiltonian:
\begin{equation}\mathcal{H}=\left.\left(\begin{array}{cccc}\omega_{c,1} & 0 & g & g\\ 0 & \omega_{c,2} & g & ge^{-i\theta}\\ g & g & \omega_{m}-i\gamma & 0\\ g & ge^{i\theta} & 0 & \omega_{m}+i\gamma\end{array}\right.\right).\end{equation}
\quad Eigenvalues are found by solving \(\mathrm{det}[K(\omega)] = 0\), where \(\mathbb{I}_4\) is the 4×4 identity matrix, and  \(K(\omega) = \omega \mathbb{I}_4 - \mathcal{H}\). We define \(\omega_c\) as the average cavity field frequency and \(\delta_c\) as the detuning between cavity fields is defined as \( \omega_c = (\omega_{c,2} + \omega_{c,1}) / 2 \), and \( \delta_c = (\omega_{c,2} - \omega_{c,1}) / 2 \).\begin{figure}[H]
  \centering
  \begin{subfigure}[b]{0.4\textwidth}
    \includegraphics[width=0.9\textwidth, height=0.7\textwidth]{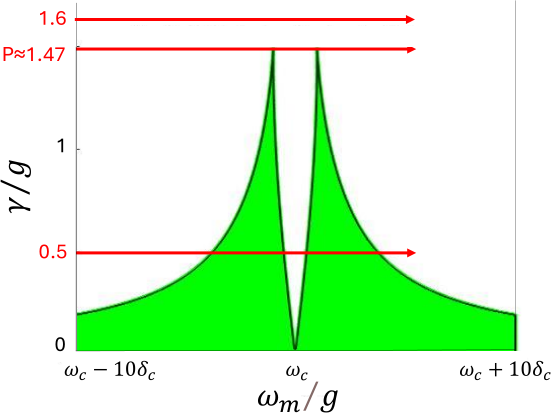}
    \caption*{}
    \label{fig:sub4}
  \end{subfigure} \vspace{-20pt}
  \caption{Pseudo-Hermitian phase diagram without coupling phase. The green region indicates the real domain; others are complex. The maximum in the green region is $\gamma / g \equiv \text{P} \approx 1.47$.}
  \label{fig:sub5}
\end{figure} 
{\textit{Result}}\ \  First, we consider the scenario without a coupling phase, where we have \(g e^{-i\theta} = g e^{i\theta} = g\). According to equation (5), we plot the pseudo-Hermitian phase diagram, as shown in Fig. 2. It is evident that the diagram is divided into two distinct regions: the green region, corresponding to the real eigenvalue domain, and the blank region corresponding to the complex eigenvalue domain. The boundary between the two regions is characterized by EPs, represented by the black line. To investigate the relationship between the pseudo-Hermitian phase transition and the energy spectrum, we selected three different strengths of non-Hermiticity in the phase diagram of Fig. 2, specifically \(\gamma/g = 0.5,  1.47\) and \(1.6\).  We first analyze the phase transitions of the system. It can be seen that for $\gamma/g = 0.5$, when the magnon frequency is relatively small, the system resides in the blank region. As the frequency increases along the red arrow, crossing the first EP, entering \begin{figure*}[t] % [t] 顶部对齐
  \centering
  \begin{subfigure}[b]{\textwidth}
    \includegraphics[width=\linewidth, height=0.75\linewidth]{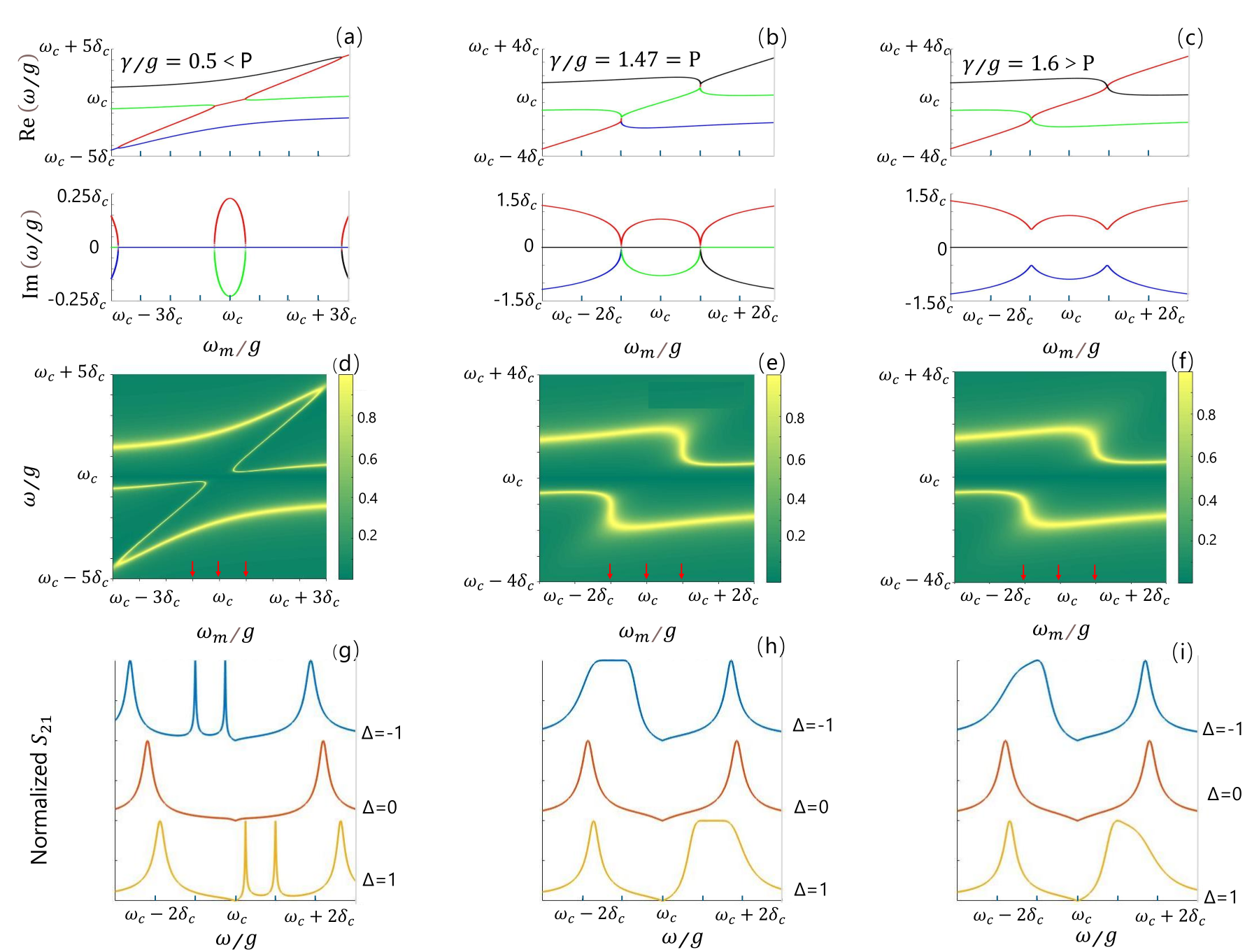}
    \caption*{}
    \label{fig:sub6}
  \end{subfigure} \vspace{-30pt}
  \caption{
    Evolution of the eigenfrequency \( \omega / g \) with the magnon frequency \( \omega_m / g \). 
    The parameters $\gamma/g$ and $\theta$ are set as follows: (a) 0.5, $0^\circ$; (b) {P}, $0^\circ$; (c) 1.6, $0^\circ$.
    The panels (d), (e), and (f) show the \( S_{21} \) spectra corresponding to panels (a), (b), and (c), respectively. 
    The normalized $S_{21}$ as a function of $\omega/g$ for different values of detuning \(\Delta\) as shown in panels (g), (h), and (i). 
    The values of \( \omega_{c,1} / g = 24 \), \( \omega_{c,2} / g = 26 \), 
    and \( \omega_{m,1} / g = \omega_{m,2} / g = \omega_m / g \).
  }
  \label{fig:sub7}
\end{figure*}the green region, and undergoing the first phase transition. It then exits the green region via the second EP, returning to the blank region, and experiencing the second phase transition. Further increasing the frequency, the system crosses the third EP, re-enters the green region, and undergoes the third phase transition,  and finally exits through the fourth EP to complete the fourth phase transition. For systems with \(\gamma / g \equiv \text{P} \approx 1.47\) (the critical value), similar four-phase-transition behavior occurs. For \(\gamma/g = 1.6 > \text{P}\), the system remains in the blank region without crossing any EPs, confirming no phase transition. Therefore, we can conclude that for systems with non-Hermiticity \(\gamma/g \leq \text{P}\) exhibit four EP-crossing phase transitions, while those with \(\gamma/g > \text{P}\) remain in the complex domain without undergoing phase transitions.\\ 
\indent We will continue to analyze the energy spectrum. In Fig. 3(a), for \(\gamma/g = 0.5 < \text{P}\approx1.47\), by examining the real parts of the eigenvalues, we can observe an attraction between the cavity mode and magnon mode in the resonance region. The corresponding imaginary parts are all zero before the level attraction occurs, indicating the absence of dissipation. When level attraction takes place, the imaginary parts begin to repel each other, signifying the emergence of dissipation. After the level attraction ends, all imaginary parts return to zero, and the dissipation vanishes. Therefore, this is a case of the level attraction induced by dissipative coupling between the magnon and cavity modes, where  the non-Hermiticity merely serving as a dissipative channel.  In the Appendix B, we calculated the transmission spectrum \( S_{21} \) of the system using input-output theory.\ucite{54} \\ 
\indent The transmission spectrum corresponding to Fig. 3(a) is shown in Fig. 3(d), which exhibits a double Z-shaped energy band. The double Z-shaped energy band observed in the transmission spectrum, along with the repulsive behavior in the imaginary part of the eigenvalues, are typical signatures of dissipative coupling. Next, we analyze the energy spectrum with the non-Hermitian strength of P, which is the largest value of the green region in Fig. 2. The results are shown in Fig. 3(b). In the real part of the eigenvalues,  we can still observe level attraction. By examining the imaginary part of the eigenvalues, we observe the same repulsive behavior as in Fig. 3(a). Therefore, the level attraction here is still induced by dissipative coupling. Since P is at a critical strength, the double Z-shaped band structure is not clearly visible in Fig. 3(e). In contrast to Fig. 3(a), Fig. 3(b) reveals a transparent window. To better illustrate the difference, we plot the normalized \( S_{21} \) as a function of $\omega/g$ for different values of detuning \(\Delta\), as shown in Fig. 3(g)-(i), where  \(\Delta = (\omega_m - \omega_c) / g\). It is clear that in Fig. 3(h), we can observe a wide plateau, corresponding to a broad characteristic frequency range. Physically, this implies that the system supports wider frequency transmission, offering advantages for applications such as tunable optical filters, high-efficiency signal processing, and quantum information transfer. Finally, we look at Fig. 3(c), where \( \gamma / g = 1.6 > \text{P} \). Observing the real part of the eigenvalues, unlike in Figs. 3(a) and 3(b), there is no longer a level attraction between cavity modes and magnon modes, but rather the two magnon modes overlap. For the imaginary part of the eigenvalues, the cavity modes exhibit zero imaginary components, indicating the absence of dissipation. In contrast, dissipation occurs solely in the magnon modes. This suggests that the system is entirely dominated by the non-Hermiticity, which is applied only to the magnon modes.

Next, we proceed to investigate the situation involving a coupling phase. When \( \theta \neq 0^\circ \), we select \( \theta = 45^\circ \) and \( \theta = 90^\circ \) to plot the phase diagram. According to Equation (5), \( g e^{i\theta} \) and \( g e^{-i\theta} \) are no longer equal. We then analyze the pseudo-Hermitian phase diagram, as shown in Fig. 4. In the phase diagram of Fig. 4(a) with \( \theta = 45^\circ \), we select two different strengths of the non-Hermiticity, and the corresponding energy spectra are shown in Fig. 5(a) and 5(b). We first analyze the phase transition of the system when \(\gamma/g = 0.5\) in Fig. 4(a). When the magnon frequency is small, the system is in the blank region. As the magnon frequency increases, the system crosses the first EP and enters the green region, undergoing the first phase transition. As the magnon frequency continues to increase, the system remains in the green region. When the magnon frequency increases further, the system crosses the second EP, enters the blank region again, and ultimately stays there, marking the second phase transition. This phase transition behaves significantly differently from the one shown in Fig. 2. Next, we analyze the phase transition of the system when \(\gamma/g = 1.4\) in Fig. 4(a). It is clear that this phase transition is similar to the one in Fig. 2, as it also undergoes four phase transitions. \begin{figure}[H]
  \centering
  \begin{subfigure}[b]{0.5\textwidth}
    \includegraphics[width=1\textwidth, height=0.454\textwidth]{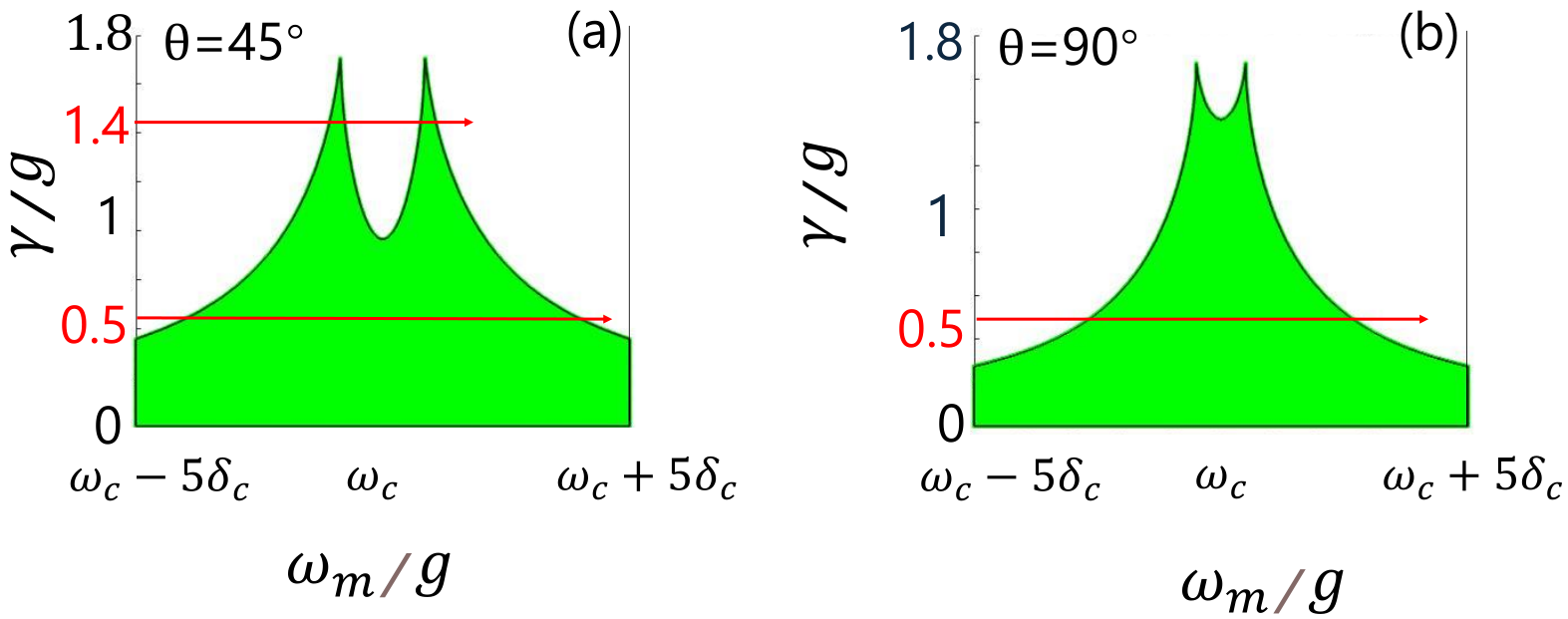}
    \caption*{}
    \label{fig:sub8}
  \end{subfigure} \vspace{-30pt}
  \caption{(a) The system phase diagram at $\theta = 45^\circ$. (b) The system phase diagram at $\theta = 90^\circ$. The strong resonance region refers to the area near \( \omega_m = \omega_c \), with the resonance center being at \( \omega_m = \omega_c \).}
  \label{fig:sub9}
\end{figure} We then analyze the corresponding energy spectrum. For Fig. 5(a), with $\gamma/g = 0.5$, although the strength of non-Hermiticity is the same as that in Fig. 3(a), the energy spectrum is dramatically different. A gap appears between two real energy bands that should have crossed when \(\theta = 0^\circ\), and they now repel each other. The imaginary parts of the eigenvalues are mutually attractive and both zero in the resonance region, similar to the \begin{figure*}[t] % [t] 顶部对齐
  \centering
  \begin{subfigure}[b]{\textwidth}
    \includegraphics[width=\linewidth, height=0.75\linewidth]{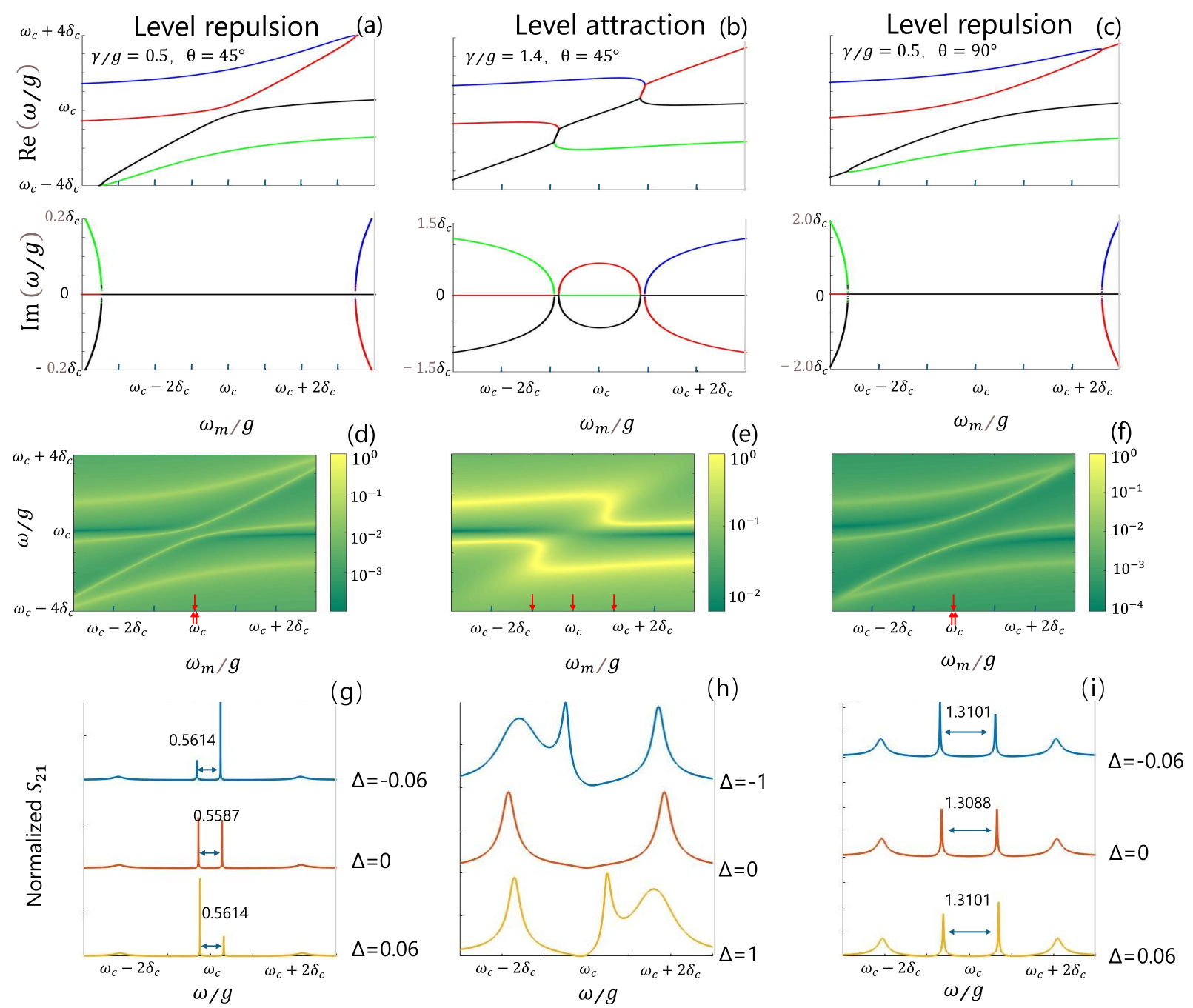}
    \caption*{}
    \label{fig:sub10}
  \end{subfigure}\vspace{-20pt}
  \caption{
    The energy spectrum of the system shows the evolution of the eigenfrequency $\omega/g$ with the magnon frequency $\omega_m/g$. The parameters $\gamma/g$ and $\theta$ are set as follows: (a) 0.5, $45^\circ$; (b) 1.4, $45^\circ$; (c) 0.5, $90^\circ$. The  energy spectrum correspond to the normalized $S_{21}$ images shown in panels (d), (e), and (f). The normalized $S_{21}$ as a function of $\omega/g$ for different values of detuning \(\Delta\) as shown in panels (g), (h), and (i).
  }
  \label{fig:sub11}
\end{figure*}coherent coupling in a Hermitian system. Fig. 5(d) is the corresponding $S_{21}$ transmission spectrum. To better highlight the details of the $S_{21}$ transmission spectrum, we plot the normalized $S_{21}$ as a function of $\omega/g$ for different values of detuning \(\Delta\). When level repulsion occurs, as shown in Figs. 5(a) and 5(c), we take \( \Delta = \pm 0.6 \) and \( \Delta = 1 \); when level attraction occurs, as shown in Fig. 5(b), we take \( \Delta = \pm 1 \) and \( \Delta = 0 \). From Fig. 5(g), we observe that when $\Delta = 0$, two equal-amplitude resonances are observed. As $\Delta$ increases, these two modes move apart, with one mode slowly decreasing in amplitude. Therefore, Fig. 5(a) is a typical example of coherent coupling\ucite{27}. Even though Fig. 5(a) and Fig. 3(a) have the same strength of non-Hermiticity, the change in the coupling phase causes the system to transition from dissipative coupling to coherent coupling. Next, we focus on the case where $\gamma/g = 1.4$. The energy spectrum in Fig. 5(b) shows attraction of the real eigenvalues and repulsion of the imaginary eigenvalues, while the corresponding $S_{21}$ transmission spectrum in Fig. 5(e) displays the double Z-shaped band structure. Therefore, Fig. 5(b) is similar to the dissipative coupling observed in Figs. 3(a) and 3(b). Thus, in the phase where \(\theta \neq 0^\circ\), the system can exhibit either coherent coupling or dissipative coupling.  The level attraction in Fig. 5(b) corresponds to a wider range of magnon frequencies compared to that in Fig. 3(a). In other words, the distance between the double Z-shaped features is larger, allowing the level attraction to occur farther from resonance. \\ 
\indent  In the phase diagram of Fig. 4(b) with \( \theta = 90^\circ \), we select \(\gamma/g = 0.5\), and the corresponding energy spectrum are shown in Fig. 5(c). We first analyze the phase transition of the system. For the phase transition when \(\gamma/g = 0.5\) in Fig. 4(b), as the magnon frequency increases, the phase transition is similar to that in Fig. 4(a) with \(\gamma/g = 0.5\). However, the difference is that, as the magnon frequency increases, the green-shaded region in Fig. 4(b)  is noticeably smaller compared to the corresponding green-shaded region in Fig. 4(a). Observing Fig. 5(c) and 5(f), we can conclude that this also corresponds to coherent coupling. However, unlike Fig.~5(a), the gap between the two repulsive energy levels in Fig.~5(c) is larger. This feature is also evident in Fig.~5(i), where the distance between the two equal-height peaks at $\Delta = 0$ is larger than that in Fig.~5(g), and this distance, which corresponds to the gap, is related to the coupling strength.\begin{figure}[H]
  \centering
  \begin{subfigure}[b]{0.5\textwidth}
    \includegraphics[width=0.9\textwidth, height=0.3\textwidth]{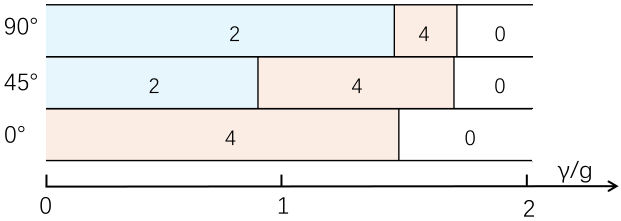}
    \caption*{}
    \label{fig:sub12}
  \end{subfigure} \vspace{-30pt}
  \caption{The distribution of phase transitions and coupling modes for different coupling phases. The blue background represents coherent coupling, and the yellow background represents dissipative coupling.}
  \label{fig:sub13}
\end{figure}  In Fig. 6, we present the distribution of the pseudo-Hermitian phase transitions and coupling modes at $\theta = 0^\circ$, $45^\circ$, and $90^\circ$. The numbers show the number of phase transitions as the magnon frequency increases for the corresponding $\gamma/g$.  According to the  above discussion, when the number of phase transitions is 2, the corresponding coupling mode is purely coherent. However, when the number of phase transitions reaches 4, the coupling mode is purely dissipative. The energy spectrum is closely connected to the phase transitions.

To show the relationship between the gap and the coupling phase, we plot the energy levels as a function of the coupling phase for $\gamma/g = 0.5$  with \( \omega_m = \omega_c \). Figs. 7(a) and 7(b) illustrate the variation of the energy levels with the coupling phase at the resonance center, revealing several distinct characteristics of our system. First, when the coupling phase takes a value less than the value at EP\(_{1}\), the gap between the two energy levels is zero. Second, when the coupling phase \begin{figure}[H]
  \centering
  \begin{subfigure}[b]{0.5\textwidth}
    \includegraphics[width=1\textwidth, height=0.4\textwidth]{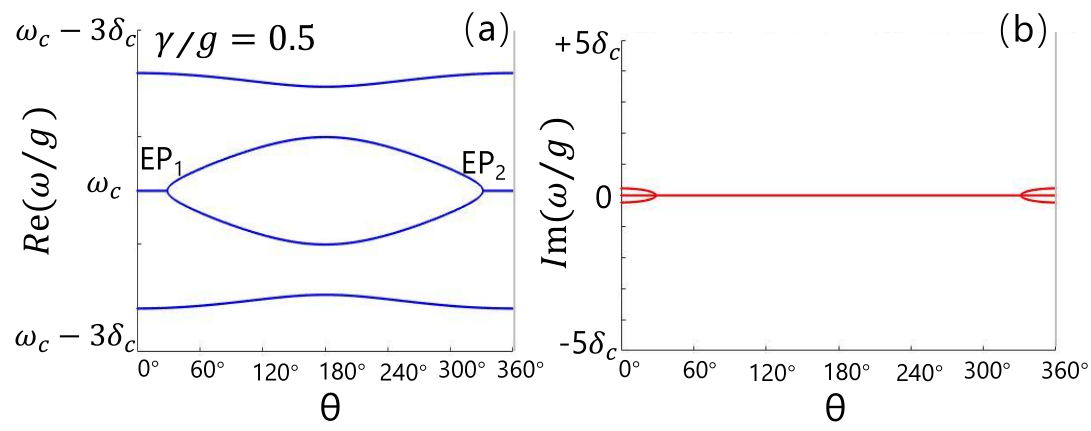}
    \caption*{}
    \label{fig:sub14}
  \end{subfigure}\vspace{-20pt}
  \caption{(a) Real parts of the eigenvalues as a function of coupling phase for $\gamma/g = 0.5$. (b) Imaginary parts of the eigenvalues. }
  \label{fig:sub15}
\end{figure} \noindent takes a value exceeds the value at EP\(_{1}\), a distinct energy gap emerges between the hybridized energy states, manifesting  as a characteristic energy level repulsion phenomenon. As the coupling phase increases, the gap between the two repulsive energy levels also increases. Third, at $\theta = 180^\circ$, the gap attains its maximum width, after which it decreases with further variation of the coupling phase. At EP$_2$, the energy gap returns to zero. Fourth, the variation of the gap with the coupling phase is clearly symmetric around \(\theta = 180^\circ\).

Finally, we simultaneously consider the coupling phase and the non-Hermiticity to study the pseudo-Hermitian phase diagram. Specifically, we set $\omega_m = \omega_c$, which corresponds to the resonance center.

Fig. 8(a) shows the pseudo-Hermitian phase diagram determined by the coupling phase and the non-Hermiticity. The green region represents the real domain, where selecting any point within this area ensures that all eigenvalues are real. The boundary of this phase diagram is also marked by EPs. Next, we calculate the pseudo-Hermitian energy band structure at  points $(\theta, \gamma/g)$ in Fig. 8(a) using equation (5). If level repulsion occurs, the energy difference $\Delta \omega$ between the two levels experiencing the repulsion at $\omega_m = \omega_c$ is recorded. If level attraction occurs, the energy difference between the two levels experiencing attraction at $\omega_m = \omega_c$ is recorded as $\Delta \omega = 0$. Similar to the system in Fig. 8(c), which is dominated by non-Hermiticity, we directly record $\Delta \omega = 0$. The results are shown in Fig. 8(b). As shown the gap reaches its maximum when $\theta = 180^\circ$. For any fixed coupling phase, as the non-Hermiticity increases, the system first experiences coherent coupling, then transitions to dissipative coupling, and ultimately becomes dominated by non-Hermiticity. \begin{figure}[H]
  \centering
  \begin{subfigure}[b]{0.5\textwidth}
    \includegraphics[width=1\textwidth, height=0.43\textwidth]{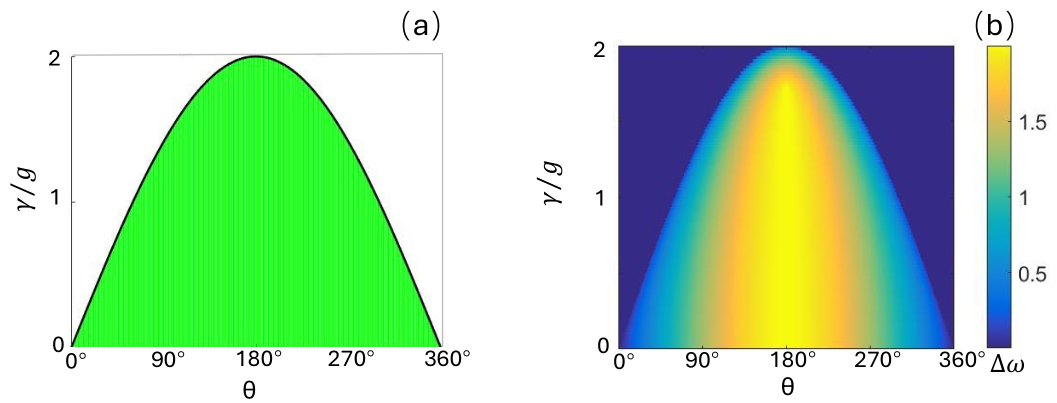}
    \caption*{}
    \label{fig:sub16}
  \end{subfigure}\vspace{-20pt}
  \caption{(a)  Phase diagram of a pseudo-Hermitian system determined by both non-Hermiticity and the coupling phase. (b)  Frequency difference between two hybridized energy levels undergoing either coherent or dissipative coupling, with color indicating the magnitude of $\Delta \omega$.}
  \label{fig:sub17}
\end{figure}
A comparison between Fig.~8(a) and Fig.~8(b) reveals a clear correspondence: the region of coherent coupling shown in Fig.~8(b) exactly matches the real domain in Fig.~8(a), while the dissipative coupling region in Fig.~8(b) coincides with the complex domain in Fig.~8(a). In other words, the transition from real to complex energy spectra, due to the breaking of pseudo-Hermiticity, directly maps to the transition from coherent to dissipative coupling modes. This correlation demonstrates that, in our constructed pseudo-Hermitian system, there is a strong connection between the pseudo-Hermitian phase transition and energy spectra.

{\textit{Conclusion}}\ \ We construct a pseudo-Hermitian system by introducing gain-loss balanced non-Hermiticity and study the relationship between its energy spectrum and phase transitions. The phase diagram shows that when pseudo-Hermiticity is broken, critical points emerge as EPs.

We first examine the case without a coupling phase. When non-Hermiticity is below the threshold $\textnormal{P}$ (Fig. 2), the system undergoes four phase transitions and crosses four EPs as the magnon frequency increases. The energy spectrum in this regime reflects dissipative coupling, where non-Hermiticity acts as a dissipative channel between the magnon and cavity modes. At the critical point $\textnormal{P}$, a broad transparent window emerges, covering a wide frequency range.

When the coupling phase is non-zero, an additional transition type emerges: a 'complex-real-complex' transition that crosses two EPs and corresponds to coherent coupling. Tuning the coupling phase allows precise control of the coherent coupling gap.

The phase serves as a control parameter for pseudo-Hermitian transitions, expanding the parameter space for such transitions. It enables the exploration of higher order EPs, exceptional surfaces, and saddle points in magnon polariton systems. Finally, the pseudo-Hermitian phase diagram, defined by non-Hermiticity and coupling phase, shows that the real domain corresponds to coherent coupling, while the complex domain corresponds to dissipative coupling. Thus, the transition between real and complex domains maps to the transition between coherent and dissipative coupling, offering new insights for controlling cavity magnon systems. Understanding this link between energy spectra and phase transitions is crucial for manipulating energy levels and coupling modes in quantum systems.

{\textit{Acknowledgments}}\ \ This work was supported by National Natural Science Foundation of China (Grant No. 12104296).

   {\centering
    \section*{\normalfont\bfseries\small Appendix A: Demonstration of Pseudo-Hermiticity}
}

If a Hamiltonian satisfies the following relation, it can be shown that the Hamiltonian is pseudo-Hermitian, where a dagger denotes the adjoint of the corresponding operator and \( \eta \) is a invertible linear operator, \begin{equation} \mathcal{H}^\dagger = \eta \mathcal{H} \eta^{-1}. \end{equation} \indent For the total Hamiltonian matrix in equation (5), we first calculate its transpose and complex conjugate. The matrix of equation (5) becomes, after taking the transpose and complex conjugate,\begin{equation} \begin{pmatrix} \omega_{c,1} & 0 & g & g \\ 0 & \omega_{c,2} & g & g e^{-i\theta} \\ g & g & \omega_m + i\gamma & 0 \\ g & g e^{i\theta} & 0 & \omega_m - i\gamma \end{pmatrix}. \end{equation}\indent Next, we define an operator \begin{equation} \beta = \begin{pmatrix} 1 & 0 & 0 & 0 \\ 0 & 0 & 1 & 0 \\ 0 & 0 & 0 & 1 \\ 0 & 1 & 0 & 0 \end{pmatrix} .\end{equation}\indent We first perform the transformation $\beta \mathcal{H} \beta^{-1}$ on the matrix in equation (5), and the result is as follows\begin{equation} \begin{pmatrix} \omega_{c,1} & g & g & 0 \\ g & \omega_{m} - i r & 0 & g \\ g & 0 & \omega_{m} + i r & g e^{i \theta} \\ 0 & g & g e^{-i \theta} & \omega_{c,2} \end{pmatrix}. \end{equation}\indent We perform another unitary transformation on the Eq. (A4), where the unitary operator is $U = e^{-i \theta c_2^{+} c_2}$, and the result after the unitary transformation is\begin{equation} \begin{pmatrix} \omega_{c,1} & g & g & 0 \\ g & \omega_{m} - i r & 0 & g e^{-i \theta} \\ g & 0 & \omega_{m} + i r & g \\ 0 & g e^{i \theta} & g & \omega_{c,2} \end{pmatrix}. \end{equation}\indent Now, by comparing Eq. (A2) and Eq. (A5), we can observe that they are the same, meaning that it satisfies $\mathcal{H}^\dagger = \eta \mathcal{H} \eta^{-1}$. Where $\eta = U \cdot \beta$.\\

   {\centering
    \section*{\normalfont\bfseries\small Appendix B: Derivation of the Transmission Spectrum}
}

First, we consider a cavity with two modes interacting with two distinct harmonic bosonic baths (environments). By incorporating the noise and dissipation terms into the Heisenberg equations for the operators, we derive the corresponding quantum Langevin equations.\begin{equation}
\begin{aligned}
\dot{c}_k &= -i \omega_{c,k} c_k - i \sum_l g_{kl} m_l - \sum_{o \in \{a,b\}} \sqrt{\beta_{o,k}^*} o^{\mathrm{in}}(t) \\
&\quad - \sum_{o \in \{a,b\}, l} \frac{\sqrt{\beta_{o,k}^* \beta_{o,l}}}{2} c_l(t),
\end{aligned}
\end{equation}  \begin{equation}
\begin{aligned}
\dot{m}_k &= -i \omega_{m,k} m_k - i \sum_l g_{lk} c_l.
\end{aligned}
\end{equation}\indent Here, \( o \in \{a,b\} \) represents the operators of two distinct harmonic bosonic baths, and \( \beta_{o,k} \) represents the leakage rate of the \( k \)-th cavity mode to the environment bath operator \( o \). The input-output relation in frequency space satisfies the following relationship: \( \tilde{o}^\mathrm{out} - \tilde{o}^\mathrm{in} = \sum_{k} \sqrt{\beta_{o,k}} \tilde{c}_k \). Here, we let all the leakage rates of a photon to the environment be \( \beta \). After calculation, the final expression for the transmission coefficient is as follows\begin{equation}
S_{21} = \frac{\left( (2 - 2 \cos(\theta)) \cdot \frac{1}{F} - 2 \frac{\omega}{g} + \sum_{i=1}^2 \frac{\omega_{c,i}}{g} \right) \cdot i \beta}{f(E,F)},
\end{equation}
\begin{equation}
\begin{aligned}
f(E,F) = & \prod_{i=1}^2 \left( \left( \frac{\omega}{g} - \frac{\omega_{c,i}}{g} \right) + i\beta  - \frac{1}{E} - \frac{1}{F} \right) \\
& - \left(i\beta - \frac{1}{E} - \frac{\cos(\theta)}{F} \right)^2 - \left( \frac{\sin(\theta)}{F} \right)^2,
\end{aligned}
\end{equation} \vspace{3mm}
\begin{equation}
E = \frac{\omega}{g} - \frac{\omega_m}{g} + \frac{i\gamma}{g}, \quad
F = \frac{\omega}{g} - \frac{\omega_m}{g} - \frac{i\gamma}{g}.
\end{equation}% 增加10mm的垂直间距

\end{multicols}
\end{document}